\documentclass[conference, onecolumn, 12pt]{IEEEtran}
\IEEEoverridecommandlockouts
\usepackage{cite}
\usepackage{amsmath,amssymb,amsfonts}
\usepackage{graphicx}
\usepackage{textcomp}
\usepackage{xcolor}
\def\BibTeX{{\rm B\kern-.05em{\sc i\kern-.025em b}\kern-.08em
    T\kern-.1667em\lower.7ex\hbox{E}\kern-.125emX}}

\usepackage{pgfplots}
\usepackage{url}
\usepackage{hyperref}
\usepackage{algorithm} 
\usepackage{algpseudocode} 
\usepackage{enumitem} 
\usepackage{xspace} 
\usepackage{float} 
\usepackage{makecell}

\pgfplotsset{compat=1.18} 

\begin{document}

\title{Towards High-Performance Multimodal Discovery: Evaluating Retrieval Efficacy and Operational Scalability in Digital Libraries \\
%{\footnotesize \textsuperscript{*}Note: Sub-titles are not captured in Xplore and should not be used}
\thanks{This research funded in part through a grant from the National Science Foundation OAC-2209872. This research reflects the views of the authors and not of the NSF.}
}
\author{
    Krishna Priya\textsuperscript{1}, 
    Sai Navya Jyesta\textsuperscript{2}, 
    Sachith Withana\textsuperscript{3}, 
    Neelesh Karthikeyan\textsuperscript{3},
    Beth Plale\textsuperscript{3} \\
    \small\textsuperscript{1}Intelligent Systems Engineering, Indiana University, Bloomington, IN USA \\
    \small\textsuperscript{2}Computer Science Dept, Indiana University, Bloomington, IN USA \\
    \small\textsuperscript{3}School of Computer and Data Sciences, University of Oregon, Eugene, OR USA \\
    \small krishnapriya051@gmail.com, sjyesta01@gmail.com, swsachith@gmail.com, neelk@uoregon.edu, plale@uoregon.edu
}

\maketitle

\begin{abstract} 
Large-scale digital repositories with million plus digitized volumes serve as vital engines for scientific and scholarly discovery. However, as collections grow, finding specific research or educational material remains a significant challenge. This difficulty is compounded by the multimodal nature of texts, that is, texts which combine dense narrative content with essential illustrative charts, annotated diagrams, and complex layouts. Historically, discovery has been supported by full-text search indexes implemented as inverted indexes (e.g., Lucene or Solr). While these systems scale well for textual strings, they are fundamentally ``blind" to the semantic nuance of visual information. 

We explore the benefit of discovery through texts that have been converted to numerical vectors. By representing page images as numerical vectors rather than OCR-extracted text, and doing so for both text and images on a page, we capture both textual and visual semantics.
The primary contribution of this research is an evaluation of the multimodal retrieval landscape, encompassing both semantic precision and architectural scalability.
Our findings demonstrate that integrating visual and textual features significantly enhances the discovery of relevant material, particularly in domains where diagrammatic content is essential for conceptual understanding. We assess retrieval efficacy, demonstrating that the system achieves sub-linear scalability. Our findings suggest that an optimized vector database configuration is the critical factor in transforming VLM-based embeddings into a production-ready, library-scale discovery solution.

\end{abstract}


\section{Introduction}

Large-scale digital repositories, such as the HathiTrust \cite{christenson2011hathitrust} with its 17 million plus digitized volumes from research libraries around the world, serve as vital engines for scientific and scholarly discovery. However, as collections grow, finding specific research or educational material remains a significant challenge. This difficulty is compounded by the multimodal nature of scientific texts, that is, texts which combine dense narrative content with essential illustrative charts, annotated diagrams, and complex layouts.

Historically, discovery has been supported by full-text search indexes implemented as inverted indexes (e.g., Lucene or Solr). While these systems scale well for textual strings, they are fundamentally ``blind" to the semantic nuance of visual information. Furthermore, by the FAIR principles (Findable, Accessible, Interoperable, {and Reusable}) \cite{wilkinson2016fair}, the principle of Findability is often the most difficult to satisfy when information is trapped within non-textual page elements.

Large Language Models (LLMs) offer a new discovery mechanism that is capable of acting on human language requests. The use of LLMs for discovery over copyrighted content has significant legal and ethical concerns because of data memorization~\cite{karamolegkou2023copyright} (quite simply, LLMs memorize data chunks). Our approach to discovery is responsive to this concern by working with text that has been converted to numerical vectors. By representing page images as numerical vectors rather than OCR-extracted text, and doing so for both text and images on a page, we capture both textual and visual semantics.

Specifically, we architect a software framework composed of a Vision-Language Model (VLM) and a vector database.  We use the framework to establish the efficacy of multimodal embeddings.  Our findings demonstrate that cosine similarity, when utilized within a shared latent space, effectively retrieves semantically and visually relevant pages from a corpus of digitized textbooks, significantly outperforming traditional distance measures like Euclidean or Manhattan distance.

However, the high-fidelity retrieval provided by VLMs introduces a new challenge: their computational viability with large digital data collections, such as the 6 billion plus digitized pages of the HathiTrust Digital Library.  While the ``intelligence" of these models is unquestioned, their performance and scalability across library-scale datasets remain under-explored. 
%As noted in prior studies, query response time is heavily dependent on the level of concurrency supported by the underlying architecture.

This paper presents a holistic evaluation that bridges both semantic discovery and operational performance. In moving to operational performance, we shift the focus from a purely experimental ``standalone" model to a robust, decoupled architecture designed to operationalize multimodal discovery. Utilizing a pipeline consisting ColPali for embedding, and the Qdrant vector database for high-performance indexing, we evaluate the system's ability to maintain high precision while achieving sub-linear scalability.

 We illustrate our study through three representative queries from our benchmark.  For each of the queries, we explain the query and then discuss the top 5 results that are retrieved under cosine similarity.  The purpose is to illustrate the kinds of similarities that are uncovered through vector embedding.
%In each case, we show that relevant material was retrieved by the system but was not included in the original gold standard, highlighting the limits of static evaluation and the value of flexible semantic retrieval in a library context.

\begin{itemize}
    \item 
\textit{“Tell me about the Bellman-Ford algorithm and give me an example”}  \\
One would expect here a description of how the algorithm works, possibly a pseudocode representation of an algorithm and either a pseudocode example or narrative form example. As pseudo-code algorithms can take up considerable space on a page, it is reasonable to expect that the results are presented across multiple consecutive pages. Hence this query, MP-14, is categorized as a multi-page query in our benchmark.

% AUTHOR CHECK: Verify the first retrieved page against the original Bellman-Ford query results; no replacement description of its technical contents has been inferred.
A detailed and well-worked example demonstrating the algorithm's practical application is the second page returned. 
%The recovered content is extremely relevant due to the conceptual connection between the application and the theoretical underpinning. This outcome offers a comprehensive understanding of the algorithm's operation by demonstrating the system's capacity to manage multi-page searches and obtain both the essential practical example and the underlying explanation.

\item
\textit{“How do state transition graphs work?”} \\ 
State transition graphs show the logical progression of a computational entity through states.  The query, MM-1, is categorized as a multi-modal query as the expected result is an image and description. While the ground truth includes one page with a labeled diagram,  a second page contains a paragraph-length textual explanation of each state and its transition. The conceptual alignment between the text and the query intent made the second page highly relevant. This result illustrates the system’s ability to connect visual and textual signals, which is particularly important in illustrated technical textbooks.

\item{\textit{“How does eventual consistency differ from strong consistency in terms of behavior and guarantees?”}} \\
This query, CP-13, is an example of a concept-driven query.  Eventual consistency is a distributed systems concept that has to do with computer processes that are geospatially distributed but have the requirement to have a consistent view of some piece of information.  Eventual consistency allows for some time to pass before the world views become consistent.

The query aims to highlight the differences between eventual consistency and strong consistency, both in terms of behavior and guarantees.  The first page retrieved provides a fundamental overview of eventual consistency, detailing its key ideas and use in distributed systems.  The second page retrieved provides a more detailed explanation of strong consistency, including its tougher assurances and the behavior it ensures in data synchronization between systems.  The theoretical foundations of both consistency models are aligned; therefore, the retrieved content is extremely relevant to the query.  %This result demonstrates the system's capacity to handle concept-driven inquiries by obtaining content that both explains and contrasts each consistency model, providing a thorough grasp of their distinctions and implications in distributed system design.
\end{itemize}

%assessment one may make of the  the contributions that could be made to discovery over large scale digital libraries through vector embeddings of multi-modal digitized texts. 
The aim of this study is to assess the viability of using vector embeddings, which reflect both text and the information in diagrammatic graphics to enhance discovery through natural language queries.
Our ultimate goal is to assess the extent to which a vector database will add to the discoverability of a digital library beyond retrieval techniques currently in place. This study is a step along that path. 

 Making sense of information in multiple forms has been explored under the banner of multimodal machine learning, a field which has been active since the late 2010 through early work by Collell et al. ~\cite{Collell_Zhang_Moens_2017}. We study how to effectively utilize available multimodal cues from diagrams for the cross-modal diagram-text retrieval task \cite{hassan2013multi}.  Specifically, we apply tools for multimodal embedding, where both text and images are represented via the same mathematical object (that is, a vector).  Multimodal embeddings represent multiple data modalities (e.g., text, graphics) in the same vector space such that similar concepts are located in vector space proximity to one another.  This means that all the multimodal data resides in a single vector database. 

Other approaches to making digital content accessible for discovery and other needs include the HathiTrust Research Center Extracted Features (EF) Dataset~\cite{9651860}. While not exclusively for discovery, the EF dataset consists of information about page structure, bags of words, and metadata drawn from OCR'ed volumes of the HathiTrust Digital Library.  The dataset has enabled unique computational analyses of the corpus and is released under a CC by 4.0 license. 

The primary contribution of this research is an evaluation of the multimodal retrieval landscape, encompassing both semantic precision and architectural scalability. To the semantic precision, we introduce a comprehensive benchmark of 75 natural language queries with manually curated ground-truth results to assess the efficacy of Vision-Language Model (VLM) embeddings. Our findings demonstrate that integrating visual and textual features significantly enhances the discovery of relevant material, particularly in domains where diagrammatic content is essential for conceptual understanding. Specifically, we establish that cosine similarity serves as the superior distance measure for maintaining semantic consistency across modalities, yielding the most reliable retrieval results.

To architectural scalability, we experimentally evaluate vector embedding and query evaluation. By evaluating the Colpali - Qdrant framework on community cloud infrastructure, we show that by optimizing indexing techniques, such as on-disk Memmap storage and binary quantization, we are able to achieve sub-linear scalability and resilient throughput under significant levels of concurrency. Taken together, these insights offer a practical, high-performance blueprint for strengthening digital repositories like HathiTrust, providing a scalable pathway for navigating the visually rich, multimodal future of scholarly discovery.

The remainder of the paper entails related work, methodology, evaluation and conclusion.

\section{Related Work}\label{sec:relatedwork}
 
The evolution of document retrieval has shifted from keyword-based indexing to dense semantic vector spaces, driven by advancements in transformer architectures and multimodal learning.

\subsection{From Keyword Search to Semantic Embedding}
Traditional document retrieval systems rely primarily on statistical techniques such as TF-IDF \cite{sparck1972statistical} and BM25 \cite{robertson1994some}, which utilize inverted indexes to match query terms against document corpora. While efficient and scalable, these methods are fundamentally limited by the ``vocabulary mismatch" problem \cite{furnas1987vocabulary} and an inability to interpret non-textual elements like diagrams or tables. The rise of Large Language Models (LLMs) and dense retrievers (e.g., DPR) \cite{karpukhin2020dense} introduced semantic search, where text is projected into a high-dimensional latent space. However, as noted by Karamolegkou et al. \cite{karamolegkou2023copyright}, these models often rely on memorization, raising significant copyright and privacy concerns in the context of research libraries.

\subsection{Multimodal VLMs and Late Interaction}
Multimodal retrieval seeks to bridge the gap between textual queries and visual content. Early Vision-Language Models (VLMs) like LayoutLM \cite{huang2022layoutlmv3} and DocFormer \cite{appalaraju2021docformer} utilised ``early interaction," where visual and textual features are fused before encoding. While powerful, these models often struggle with fine-grained alignment between specific query tokens and localised image patches.

The introduction of ColBERT \cite{khattab2020colbert} and its multimodal successor, ColPali \cite{faysse2024colpali}, introduced the ``late interaction" paradigm. By representing documents as a sequence of multiple vectors (multi-vector representations), these models allow for token-to-patch alignment at the scoring stage. This approach is particularly effective for textbooks where information is spatially distributed across diagrams and text. This study builds on the efficacy of ColPali but extends the research beyond simple retrieval quality to address the systemic overhead of managing such high-cardinality multi-vector data.

\subsection{Vector Database Evolution and Optimization}
As retrieval systems transition from laboratory experiments to production-grade tools, the role of the vector database has become central. Specialized systems such as Milvus, Weaviate, and Qdrant \cite{xie2023brief} utilize Hierarchical Navigable Small World (HNSW) \cite{malkov2018efficient} graphs to achieve approximate nearest neighbor (ANN) search.

Recent research in High-Performance Computing (HPC) environments has highlighted that vector database performance is frequently CPU-bound, particularly during the conversion of embeddings into database-specific batch objects. Furthermore, the storage of multi-vectors (often exceeding 1,000 vectors per page) poses a significant memory challenge. Our work explores the use of on-disk Memmap storage and binary quantization within Qdrant to overcome these memory bottlenecks, providing a unique performance characterization that is currently missing from the literature on VLM-powered digital libraries. 
%The challenge of achieving FAIR principles in large-scale repositories like HathiTrust has been extensively explored, particularly regarding the `Findability' of non-textual elements trapped in digitized volumes. While the HathiTrust Research Center’s Extracted Features (EF) 2.0 dataset provides essential structural and metadata cues , it remains fundamentally limited by the vocabulary mismatch of OCR-derived text. Our work extends recent JCDL explorations into image-based multimodal retrieval by operationalizing the late interaction paradigm to resolve fine-grained spatial relationships between diagrams and narrative text.

\section{Methodology}\label{methodology}

\begin{figure}[htbp]
    \centering
    \includegraphics[width=11cm]{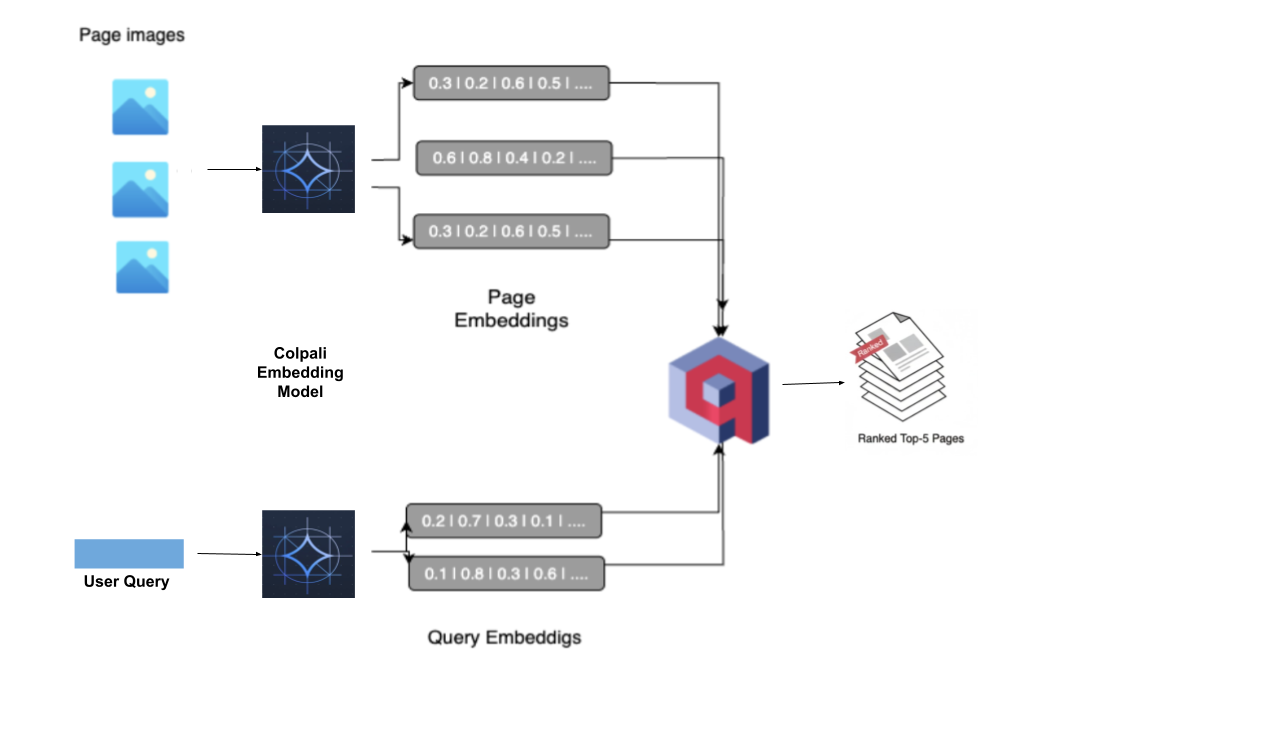}
    \caption{Architectural Schematic}
    \label{fig:workflowDiagram}
\end{figure}

The software framework (architecture) is composed of a Vision-Language Model (VLM) and a vector database, as illustrated in Fig. \ref{fig:workflowDiagram}. Across the top of the figure shows page images passed through a Vision-Language Model (VLM). What results are a set of vector embeddings that are then stored to a vector database. This would be done at once to "vectorize" a digital library collection. A VLM approach eliminates the need for traditional, error-prone preprocessing pipelines involving Optical Character Recognition (OCR), layout detection, and manual chunking. The vector database supports multi-vector indexing meaning that a data object is represented by multiple vectors rather than as a single vector.

Along the bottom of the figure is the search query.  This too needs to be embedded so is passed through the VLM to obtain a vector embedding.  It occurs later in time and can be done on a per-query basis, that is, as a query is issued. The vectorized query is stored to the vector database and from there it is "executed" to find all relevant pages that have a minimal distance from the query using a pre-defined distance measure. The top relevant pages are returned in order of distance. What constitutes "top" is set by parameter, usually 5 or 10.  Results are selected based on similarity scores (distance measure) between the query and document embeddings, computed using one of four distance functions (Cosine, Dot Product, Euclidean, Manhattan).  

This approach works because {ColPali} supports a ColBERT-style late interaction search, allowing the query to be issued over a set of existing embeddings. 
% Each query produces a ranked list of pages, from which the top-k results are returned for evaluation. 

In this paper, we study the architecture for its operational performance and for its ability to expose new information about document relationships that are made possible by a vector distance determination of relationship.  The framework decoupled design additionally allows us to independently evaluate embeddings generation and the vector retrieval processes.   We describe each component in more detail below. 

We bring a dataset of 3,612 image-based textbook pages and a small recipe book that is rich with pictures. The textbooks are five popular undergraduate textbooks in computer science where we have rights to the digital copy. The text books are in operating systems, programming languages, computer systems, and AI. 

\subsection{Vision-Language Embedding: ColPali and PaliGemma-3B}

%\begin{itemize}   \item %\textit{Operating System Concepts} (10th Edition) –     operating systems concepts and images of process management, memory allocation, and file systems.
% \item %\textit{Artificial Intelligence: A Modern Approach} (4th Edition) – state-of-the-art 
 %   AI, machine learning, planning, and search abstractions, processes.
  %  \item %\textit{Introduction to Automata Theory, Languages, and Computation} –  formal languages, automata, grammar, and Turing machines.
%  \item %\textit{Principles of Computer System Design} –    layered architecture, abstraction, system interfaces, and fault tolerance.
%   \item %\textit{Cook’s Illustrated Annual 2019 Recipes} –  food recipes. Included as a contrasting domain to evaluate model generalization across non-technical, visually structured content.
%\end{itemize}

Each volume is originally in the form of a single document in PDF format. 
%and was preprocessed through the following steps to create a consistent input format for our retrieval system:
%Each page (PDF) is split into individual pages using \texttt{PyMuPDF}, ensuring that page boundaries and formatting were preserved.
Using \texttt{PyMuPDF}, the full volume PDF document is broken into individual pages where each page is rendered as an image with a DPI of 300, which is considered a high resolution image standard.  DPI of 300 balances readability of textual and visual elements with computational efficiency. The resulting image is stored as a PNG image; PNG ensures lossless compression and layout fidelity, which is essential for downstream visual embedding.

% AUTHOR CHECK: Confirm the exact checkpoint and the image-token, text-token, and padding composition of the reported 1030-vector sequence.

{The page images are encoded by ColPali using its PaliGemma-3B backbone, which combines a SigLIP vision encoder with a Gemma-2B language model. ColPali projects the resulting representations into a shared 128-dimensional embedding space, producing a sequence of vectors for each page that captures its textual, visual, and structural content.}

The key to our scalable approach is a technique pioneered in 2020 by the ColBERT information retrieval model ~{\cite{khattab2020colbert}} called \textit{Late Interaction Scoring (LI)}. LI is a retrieval architectural design in which query and document representations are processed independently and only interact at the final stages (when scoring is applied to identify the most relevant pages). Unlike early interaction models that fuse query and document embeddings before or during encoding (e.g., through cross-attention layers), LI models maintain separate encoding pipelines, enabling efficient storage and reuse of document embeddings. The core idea is to represent both queries and documents as sequences of vectors (e.g., token embeddings for text or patch embeddings for images), and compute fine-grained relevance scores through token-to-token or token-to-patch similarity comparisons.

Specifically, in ColBERT, and later ColPali, late interaction works as follows: a document is scored by taking the maximum similarity between each query vector and all document vectors, and then aggregating these maxima into a single relevance score. Applied to our system, each textbook page is represented by a fixed set of visual patch embeddings, while each natural language query is transformed into a sequence of dense token embeddings. At retrieval time, each query token interacts with all document patches via a configurable similarity metric (e.g., cosine, dot product), and a late aggregation operation (e.g., max-sum pooling) produces the final document relevance score.

The late interaction approach enables token-level alignment between query concepts and fine-grained visual features of the page (e.g., specific diagrams, formulas, or text blocks), making it well-suited for visually dense documents like textbooks. Moreover, late interaction scoring allows us to isolate and evaluate the effect of different similarity functions (cosine, dot product, Euclidean, Manhattan) within a controlled and interpretable retrieval {pipeline.}
% EDIT: Removed the repeated backbone, preprocessing, and projection description; these details are consolidated above.

A critical requirement for successful multimodal retrieval is the cross-modal alignment of query and document embeddings within a shared latent space. In this pipeline, query vectorization is intrinsically coupled with the ColPali model to ensure semantic compatibility across modalities. As noted in foundational multimodal literature \cite{radford2021learningtransferablevisualmodels}, maintaining a shared manifold is essential for zero-shot retrieval tasks. While it is theoretically possible to utilize disparate tools for query and image vectorization, such an approach would necessitate a complex transformation matrix or a shared contrastive training objective (e.g., CLIP-style pre-training) to map textual queries into the same high-dimensional manifold as the page images.

In our implementation, we mitigate the risk of ``manifold misalignment" by utilizing ColPali’s internal language model backbone to vectorize queries. This ensures that the resulting textual token embeddings are natively comparable to the visual patch embeddings. This architectural coupling allows for the direct application of the MaxSim operator, as both the query and the document reside in a bi-modal projection space where distance represents true semantic similarity rather than just modal correlation. This alignment is what enables the model to resolve ``fine-grained" queries, such as linking a technical term in the query to a specific annotated component in a complex architectural diagram.

\subsection{Indexing Layer: Qdrant Vector Database}
The core search and scoring engine is implemented using the Qdrant Vector Database. All page embeddings are stored (indexed) to a vector database. We selected the \textit{Qdrant Vector Database}~\cite{qdrant2024} as it is high-performance, open source, and allows for multi-vector indexing which means each document (\textit{i.e.}, page) can be stored as a sequence of vectors, rather than a single aggregated representation.  Qdrant supports executing ColBERT-style late interaction scoring directly within the database engine.

To optimize the system for library-scale datasets on resource-constrained hardware, we implemented three specific database configurations:

\begin{itemize}
    \item \textbf{HNSW Indexing:} We utilize a Hierarchical Navigable Small World (HNSW) index to achieve sub-linear search complexity across the $>4$ million patch vectors in our corpus.
    \item \textbf{On-Disk Memmap Storage:} We utilize Memmap storage to persist vector data directly on the disk. This allows the system to scale to datasets that exceed available RAM, with the operating system managing page caching for frequently accessed vectors.
    \item \textbf{Binary Quantization:} To reduce I/O bandwidth and increase throughput, we enabled binary quantization. This compresses the 128-dimensional floating-point vectors into binary representations for the initial candidate search phase, significantly accelerating the MaxSim calculation without sacrificing semantic precision.
\end{itemize}

The design is guided by the need for rigorous experimental control, where all aspects of the retrieval pipeline are held constant across all experiments, including embedding model, dataset, indexing method, query formulation, and evaluation protocol. The only variable that can be altered between runs is the choice of similarity function used during search of the vector database. This strict isolation allowed us to attribute observed performance variations solely to the behavior of the distance metric in high-dimensional semantic space.  This is an important factor as one of our evaluations is to systematically assess how different vector distance metrics (i.e., Cosine similarity, Dot Product, Euclidean distance, and Manhattan distance) affect the quality and relevance of retrieved results.

\subsection{Query-based Image Retrieval} 

Queries are issued to each configuration independently, and retrieved results compared against a manually curated ground truth using standard information retrieval metrics. To evaluate the impact of different similarity measures on retrieval performance, we configure four distinct vector distance functions within the vector database. 
Each distance function is configured for use in a separate Qdrant collection having identical content and configuration to the others except for the distance function.  The distance functions are shown here.

\begin{itemize}

    \item \textbf{Cosine Similarity:}  
    Cosine similarity measures the cosine of the angle between two vectors. It is invariant to vector magnitude, making it well-suited for high-dimensional semantic representations such as text or image embeddings. It emphasizes direction over magnitude.

    \[
    \text{CosineSim}(\mathbf{A}, \mathbf{B}) = \frac{\mathbf{A} \cdot \mathbf{B}}{\|\mathbf{A}\| \|\mathbf{B}\|}
    \]

    \item \textbf{Dot Product:}  
    Dot Product computes the raw inner product of two vectors. Unlike cosine similarity, it is sensitive to the magnitude of vectors. It can be useful when the scale of embeddings carries semantic meaning, though it may bias results toward longer vectors.

    \[
    \text{DotProduct}(\mathbf{A}, \mathbf{B}) = \sum_{i=1}^{n} A_i \cdot B_i
    \]

    \item \textbf{Euclidean Distance (L2):}  
    Euclidean distance calculates the straight-line distance between two points in vector space. It penalizes both magnitude and direction differences. It is effective in low-dimensional or dense feature spaces but may be less robust in high-dimensional sparse embeddings.

    \[
    d_{\text{Euclidean}}(\mathbf{A}, \mathbf{B}) = \sqrt{\sum_{i=1}^{n} (A_i - B_i)^2}
    \]

    \item \textbf{Manhattan Distance (L1):}  
    Manhattan distance, also known as city-block distance, sums the absolute differences between vector components. It is less sensitive to outliers than Euclidean distance and can be more appropriate for sparse vectors.

    \[
    d_{\text{Manhattan}}(\mathbf{A}, \mathbf{B}) = \sum_{i=1}^{n} |A_i - B_i|
    \]

\end{itemize}

All other components including embeddings, the $Top-5$ value, retrieval logic, and evaluation metrics are held constant to ensure that the effect of the similarity function can be analyzed in isolation.

With the page embeddings indexed in the vector database, queries from the benchmark are issued against the database. 
After applying the similarity function, the system scores the retrieved pages based on how closely they match the query. This step helps identify the most relevant pages from the initial set of results. The system ranks the pages by comparing the query with each one and determining their relevance.  For each query, returned are the top relevant pages ($Top-k$).  We configure the database to return 25 results, then use a local algorithm to identify the top 5. The local algorithm is simple; it merely removes duplicates where they occur.

Max-sum pooling is used to look at each part of the page and select the part that matches the query the best. The best matches are combined to get a final score for the page. This ensures the system focuses on the most relevant parts of each page.

Once the initial scoring is done, the next step is re-ranking, which fine-tunes the results. Here the order of documents is adjusted based on how highly relevant they are. This helps make sure that the pages that are most likely to be useful to the user are shown first.

\section{Retrieval Efficacy Evaluation} 

We conduct a multi-dimensional assessment of the proposed discovery pipeline, evaluating its effectiveness across two primary axes: \textit{retrieval efficacy} (the qualitative accuracy of multimodal discovery) and \textit{system performance} (the operational efficiency and scalability of the underlying architecture).
The efficacy evaluation is designed to quantify the model's capacity for resolving complex multimodal relationships and to identify the optimal distance metric for high-cardinality, VLM-generated multi-vectors. 

To ensure that observed performance variations are attributable solely to the behavior of the distance metric within the high-dimensional semantic space, we maintain strict experimental control. All variables including the embedding model, index structure, and query formulation are held constant across all test runs.

\subsection{Experimental Taxonomy}
We conduct four distinct experiments to evaluate the system's robustness across varied retrieval contexts:
\begin{itemize}
    \item \textbf{Experiment I (Full Benchmark):} A comprehensive evaluation across the entire 75-query set to identify the primary distance metric.
    \item \textbf{Experiment II (Multi-page Queries):} A targeted study of the 14 multi-page queries to assess performance on spatially separated content (e.g., algorithms spanning consecutive pages).
    \item \textbf{Experiment III (Conceptual Query Subset):} An evaluation of the 13 conceptual queries to measure the model's higher-level reasoning and cross-domain linking.
    \item \textbf{Experiment IV (Unbounded Analysis):} {A study of relevant but unlabeled retrievals}, examining high-scoring results that fall outside the manually curated Top-5 ground truth.
\end{itemize}

\subsection{Query Benchmark and Taxonomy}
To cover a broad spectrum of retrieval challenges, we designed a benchmark of 75 natural language queries. These are organized into seven categories, as detailed in Table~\ref{tab:query-benchmark2}, spanning from basic textual definitions to complex, multimodal reasoning tasks. 

Each query was manually validated through a two-stage process to ensure that: (1) a verifiable answer exists within the 3,612-page corpus, and (2) the phrasing is specific enough to avoid ambiguity while remaining representative of real-world educational inquiries. Table~\ref{tab:query-benchmark2} summarizes each query type

%    \begin{itemize}
%\item{Visual queries} example “Show a figure of process synchronization and its different states.” 
%\item{Textual queries} example “What conditions lead to a deadlock” assess semantic understanding using traditional natural language search.
%\item{Multi-modal queries} example: “How does the reduction process construct a machine that distinguishes whether another machine accepts a given input?”
%\item{Tabular queries} example: “What are the I/O port address ranges assigned to different devices in a typical PC configuration?” 
%\item{Numerical queries} example “What is the cost difference between the path found by the weighted A search and the path found by standard A in terms of percentage?” 
%\item{Multi-page queries} example “What are the key design trade-offs in distributed systems?” 
%\item{Conceptual queries} example is “What is preemptive scheduling and how does it differ from non-preemptive scheduling?” 
%\end{itemize}

%when visualizing results, take 1 representative query from each category.  Set "top" to infinity. plot with distance on Y axis and data points for each response.  There will be 5 tables as result.  Drop these into paper and give the query that you used, and \# responses. 

\begin{table}[htbp]
    \caption{Query Benchmark Summarization}
       \label{tab:query-benchmark2}
    \centering
    \renewcommand{\arraystretch}{1.25}
    \setlength{\tabcolsep}{2pt}
    \resizebox{0.95\textwidth}{!}{
        \begin{tabular}{|l|l|l|c|}
            \hline
            \textbf{Query Category} & \textbf{Description} & \textbf{Example} & \textbf{Count} \\
            \hline
            Visual & Diagrams, flowcharts, or illustrations conveying conceptual content & Show figure of process synchronization with different states & 9 \\
            Textual & Asks for definitions, explanations, or semantic details in natural language. & What conditions lead to a deadlock” assess semantic understanding using traditional natural language search & 14 \\
            Multi-Modal & Requires understanding of both visual and textual content, such as labeled figures with descriptions. & How does the reduction process construct a machine that distinguishes whether another machine accepts a given input? & 10 \\
            Tabular & Involves structured data such as parameter tables, configuration matrices, or protocol listings & What are the I/O port address ranges assigned to different devices in a typical PC configuration? & 7 \\
            Numerical & Retrieve numeric values like thresholds, performance metrics, or quantitative comparisons & What is cost difference between path found by weighted A search and that found by standard A in terms of percentage? & 8 \\
            Multi-Page & Requires retrieval of conceptually linked content spread across multiple pages (e.g., algorithm + figure) & What are the key design trade-offs in distributed systems? & 14 \\
            Conceptual & Focuses on abstraction and higher-level reasoning, links visual and textual domains & What is preemptive scheduling and how does it differ from non-preemptive scheduling?”  & 13 \\
            \hline
            \textbf{Total} & & & \textbf{75} \\
            \hline
        \end{tabular}
    }
\end{table}

%\color{purple}Beth: how many queries in each category?  Overall discuss why 65 queries, and why this many in one category, this many in another, etc.)\color{black}

%\red{not clear how technical 5 from semantic-10 differ from those queries in baseline-75}\black{}
%navya’s added text
%\paragraph{\textbf{Semantic-10 Query Set}} The Semantic-10 benchmark has 10 queries, five \textit{technical} and five \textit{conceptual} questions. 

The benchmark includes factual and technical queries to test the system’s ability to retrieve content with high semantic specificity. Some will include a mix of visual or layout-based cues.  Other queries isolate domain-specific terminology and target precise definitions or algorithmic explanations without relying on structural hints (cues). The latter are well suited to assess whether the system can resolve fine-grained semantic meaning from embeddings alone, especially when the phrasing of the query does not directly match the surface text.

%Non-conceptual queries are structured to target well-defined and precise knowledge, such as “On which page is Chomsky Normal Form explained?”. Responses require high specificity and semantic matching. 
Conceptual queries, on the other hand, focus on abstraction and require the model to link higher-level reasoning across modalities. For instance, the two questions \textit{“Which page explains the intuition behind eventual consistency?”} and \textit{“Where is deadlock recovery illustrated with both text and diagram?”} test the model's ability to interpret questions that span both visual and textual domains.

\subsection{Query Set Validation} Each query is manually validated through a two-stage process. First, we manually verify that at least one relevant answer exists within the dataset of 3,612 textbook pages. Second, we refine query phrasing for clarity and specificity, ensuring that questions are not overly ambiguous or under-specified. Queries that fail these criteria are revised or excluded. This process ensures that all queries are clearly understandable and with verifiable answers in the corpus.

To maintain experimental consistency, every query is applied uniformly across all retrieval pipelines. The only changing variable across experiments, as mentioned earlier, is the similarity function used during vector comparison in Qdrant: Cosine, Dot Product, Euclidean, or Manhattan. All other components, including the embedding model (PaliGemma-3B), retrieval configuration, and evaluation metrics, are held constant. 
%This control allowed us to isolate and study the influence of each distance metric on the quality and contextual relevance of retrievals.

This structured and controlled approach to query design additionally allows for detailed performance analysis across semantic, structural, and visual dimensions. By anchoring the evaluation in both benchmark and semantic query sets, we ensure robustness, repeatability, and insight into metric-specific retrieval behavior across diverse query types.

\subsection{Evaluation Metrics}
%To evaluate the effectiveness of different similarity metrics on semantic document retrieval, we designed a comprehensive evaluation framework combining both quantitative metrics and qualitative analysis. Our goal was to understand not just which metric performs best numerically, but also which one retrieves the most contextually relevant content for educational use cases. All experiments were conducted on a fixed dataset of 3,612 page images using a consistent retrieval pipeline, with only the similarity metric varying across runs.

%We first compare the different distance measures. The similarity measure comparison is carried out on the Baseline-75 benchmark. Each query has a predetermined gold standard result that is manually curated. The purpose of the evaluation is to determine the best similarity measure on the Baseline-75 query benchmark.  

%We evaluate retrieval effectiveness using standard information retrieval metrics: \begin{itemize}
%    \item \textbf{Precision@3}: Proportion of relevant documents among the top-3 retrieved.
 %   \item \textbf{Recall@3}: Proportion of relevant documents successfully retrieved in the top-3.
  %  \item \textbf{F1 Score@3}: Harmonic mean of precision and recall.
   % \item \textbf{Average Precision (AP)}: Measures both relevance and ranking.
    %\item \textbf{Reciprocal Rank (RR)}: Inverse of the position of the first relevant result.
%\end{itemize}

We use a set of \textit{quantitative metrics} commonly employed in information retrieval tasks:

\begin{itemize}
    \item \textbf{Precision@5:} The fraction of the $top-5$ retrieved pages that are relevant (with respect to the ground truth).
    \item \textbf{Recall@5:} The proportion of all known relevant pages that appear in the $top-5$ results.
    \item \textbf{F1 Score@5:} The harmonic mean of Precision@5 and Recall@5, which provides a balanced view of both relevance and completeness.
    \item \textbf{Average Precision (AP):} A measure that accounts for both the relevance and ranking order of retrieved pages, assigning higher weight to relevant pages retrieved earlier.
    \item \textbf{Mean Reciprocal Rank (MRR):} The reciprocal of the rank at which the first relevant page is retrieved, averaged across all queries. This rewards systems that return highly relevant content earlier in the result list.
\end{itemize}

We adopt a top five retrieval evaluation framework to reflect both real-world usage patterns and experimental constraints. Users interacting with educational retrieval systems typically focus on the first few results, expecting high-precision responses without needing to sift through large result sets. Thus, evaluating only the top 5 retrieved pages provides a realistic and user-centered measure of performance. Methodologically, it ensures consistent comparison across similarity functions such as cosine and dot product by minimizing the impact of their differing score distributions. The limited result set also makes it feasible to conduct detailed manual validation across the benchmark.

\subsection{Results}

\paragraph{Experiment I: Full Benchmark}
In this first experiment, we evaluate similarity functions using the full benchmark. The retrieval process utilizes \textit{late interaction} which processes queries and documents separately until the final stages of the retrieval process. This approach, popularized by the ColBERT ranking model, allows for more efficient and precise retrieval~\cite{khattab2020colbert}. 

Results are determined by summing up the resulting scores for all query tokens to yield a final relevance score per page. The top five pages with the highest scores are selected as the retrieved results. This Top-k selection is performed independently for each metric and query. The use of a fixed $k$-value ensures that evaluation metrics remain comparable across different scoring schemes, regardless of the absolute magnitude or range of scores produced by each similarity function.

The average metric scores across the full benchmark can be found in Table~\ref{tab:metric_comparison}.
%\color{purple}We want two versions of this table, one for  baseline-65 query set and one that isolates just the multi-page queries. Take the concept-10 queries out of these results.  \color{black}
While Dot Product comparably performs against Cosine in raw metrics, deeper inspection reveals its tendency to retrieve results biased by vector magnitude, occasionally surfacing less relevant but numerically high-scoring pages. In contrast, Cosine similarity offers a more consistent semantic signal. Cosine similarity is particularly effective for 1) abstract or paraphrased queries, 2) queries involving diagrams or tabular layouts, and 3) visually dense pages with minimal text.

\begin{table}[H]
\centering
\caption{Benchmark Comparison Across Similarity Functions}
\label{tab:metric_comparison}
\begin{tabular}{|l|c|c|c|c|}
\hline
\textbf{Metric}   & \textbf{Cosine} & \textbf{Dot} & \textbf{Euclidean} & \textbf{Manh} \\ \hline
Prec@5            & 0.514        & 0.496      & 0.234           & 0.418      \\ \hline
Recall@5          & 0.281        & 0.272      & 0.125          & 0.218      \\ \hline
F1@5              & 0.353       & 0.341      & 0.160          & 0.284      \\ \hline
AP          & 0.238        & 0.233     & 0.098       & 0.184      \\ \hline
MRR        & 0.801      & 0.787      & 0.418           & 0.738      \\ \hline
\end{tabular}
\end{table}

Euclidean and Manhattan distances perform significantly worse. Their sensitivity to both vector magnitude and dimensionality make them poorly suited for high-dimensional embeddings, often returning semantically incoherent results or textually similar but irrelevant pages.

We dive in on one specific query:
\textit{“How do state transition graphs work?”} from our use case queries in the use cases. This is an example of a multi-modal query because state transition graphs are frequently presented both diagrammatically and described in narrative form. 
%While the gold standard included one page with the state graph diagram, our system retrieved a secondary page containing a paragraph-length textual explanation of each state transition. Though it lacked the diagram itself, the conceptual alignment between the text and the query intent made it highly relevant. \color{red}can we modify last sentence to this\color{black}
Within the Top-5 results is contained both the diagram and a textual explanation, with the diagram illustrating the state transition graph and the text providing a detailed explanation of each state transition. This combination of visual and textual content ensured a comprehensive response, fully aligning with the query's multi-modal intent.

No distance measure achieved strong precision accuracy, indicating that future work could explore page-level re-ranking models or context-aware retrieval strategies.

\paragraph{Experiment II: Multi-page Queries}

In this second experiment we use a subset of the full benchmark, the 14 multi-page queries, to see if the results change our conclusion with respect to the best distance measure.

Multi-page queries target a response that likely spans more than one page, for instance, a software algorithm that spans multiple pages, or a graph on one page and related text on another page.

\begin{table}[H]
\centering
\caption{Multipage Query Subset Comparison}
\label{tab:metric_comparison_multi-page}
\begin{tabular}{|l|c|c|c|c|}
\hline
\textbf{Metric}   & \textbf{Cosine} & \textbf{Dot} & \textbf{Euclidean} & \textbf{Manh} \\ \hline
Prec@5            & 0.457        & 0.442      & 0.157          & 0.342    \\ \hline
Recall@5          & 0.239        & 0.231      & 0.069           & 0.176      \\ \hline
F1@5              & 0.306       & 0.296      & 0.093           & 0.232      \\ \hline
AP          & 0.176        & 0.169      & 0.037           & 0.138      \\ \hline
MRR        & 0.690       & 0.654      & 0.246          & 0.571      \\ \hline
\end{tabular}
\end{table}

Results from our second experiment, see Table~\ref{tab:metric_comparison_multi-page}, shows the average metric scores across the multi-page queries. One can observe that cosine outperforms the other distance measures across the metrics.  Note the lower mean reciprocal rank (MRR) score compared to the full benchmark which holds across all of the distance measures.

%One can conclude a several things from the results. First, there is higher semantic continuity across retrieved pages, second, there is better relative ordering in multi-page contexts, and third, there are fewer out-of-context or redundant results. 

The query from our use cases, \textit{“Tell me about Bellman-Ford algorithm and give an example?”} requires both the theoretical explanation and a worked example. As algorithms can take up considerable space on a page, it is reasonable to expect that the results are presented across two or more consecutive pages. Hence this query is an example of a multi-page query. 

Analyzing the results from this specific query, our ground truth labeled only the initial page with the algorithm description. However, our system retrieved a second page (not labeled in the ground truth) that included a diagram illustrating the step-by-step relaxation process on a sample graph. Human and GPT-4 validation confirmed its relevance. 

\paragraph{Experiment III: Conceptual Query Subset}
In our third experiment we use the 13 queries from the Conceptual subset of the benchmark.  As with Table~\ref{tab:metric_comparison_multi-page}, MRR is slightly better but still low compared to the full benchmark. Precision of the Top 5 and Average Precision are slightly better than for the multi-page subset but still below the overall benchmark. 

%\color{red}insert table here\color{black}
\begin{table}[H]
\centering
\caption{Conceptual Query Subset Comparison}
\label{tab:metric_comparison_conceptual}
\begin{tabular}{|l|c|c|c|c|}
\hline
\textbf{Metric}   & \textbf{Cosine} & \textbf{Dot} & \textbf{Euclidean} & \textbf{Manh} \\ \hline
Prec@5            & 0.461        & 0.400      & 0.184          & 0.323      \\ \hline
Recall@5          & 0.226        & 0.198      & 0.087           & 0.157     \\ \hline
F1@5              & 0.303        & 0.264      & 0.118           & 0.210      \\ \hline
AP          & 0.182        & 0.164      & 0.059           & 0.126      \\ \hline
MRR        & 0.692       & 0.653      & 0.358           & 0.673      \\ \hline
\end{tabular}
\end{table}

\paragraph{Experiment IV: Unbounded Analysis}
Our final experiment studies relevant but unlabeled retrievals. This is where we look beyond the Top-5 results. While the Top-5 results provides us with a focused and comparable assessment, it necessarily leaves out information about the tail of the similarity distribution. To address this limitation, we carry out an unbounded retrieval analysis, where we retrieve all indexed documents and plot similarity scores across the full ranked list.

This experiment again uses the full benchmark to explore the retrieval system’s behavior in response to challenging queries especially those that may surface relevant content overlooked by the ground truth. By expanding our evaluation beyond Top-5 and beyond rigid labels, we aim to learn how well the system generalizes in less constrained settings.

\begin{figure}[htbp]    % Changed to figure* for full width
    \centering
    \includegraphics[width=7cm]{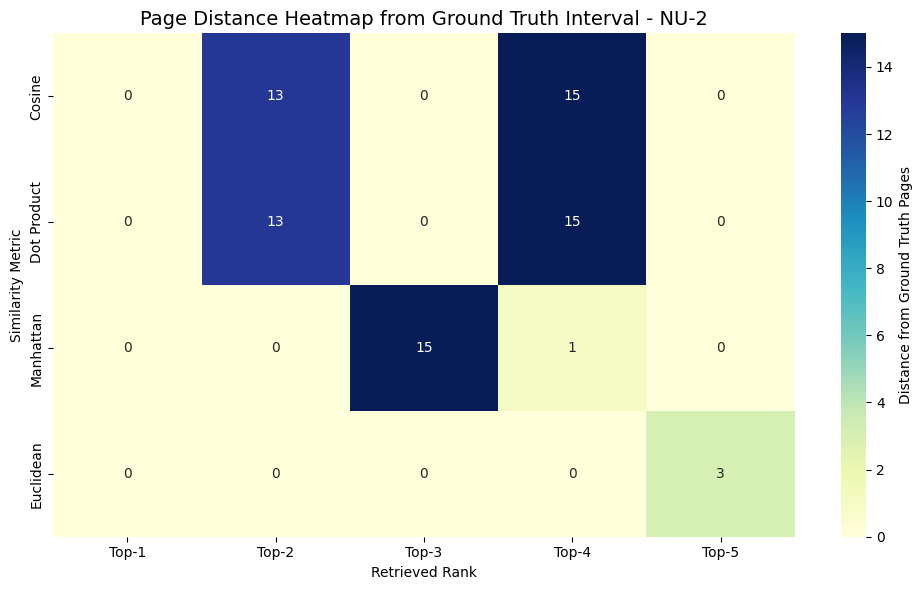}
    \caption{Page distance for numerical query NU-2}
    \label{fig:heatmap}
\end{figure}

%\color{blue}describe figures (have more than one).\color{black}

We conduct this evaluation using both Cosine similarity and Dot Product which are the top two performing distance functions from our earlier experiments. For each query, we identify pages returned outside the Top-5 that exhibit strong semantic alignment with the query but are not present in the original ground truth set. %We classify these as \textit{relevant-but-unlabeled}. 
To assess their value, we:

\begin{itemize}
    \item Measure similarity scores between the query and the retrieved page embeddings using both Cosine and Dot Product.
    \item Perform manual annotation and GPT-4 based validation to confirm contextual alignment.
\end{itemize}

The second step exposes limitations in rigid evaluation and highlighted the strengths of angular-based metrics like Cosine Similarity, as well as the density-driven retrieval capacity of Dot Product. Both metrics retrieve high-scoring pages that are semantically aligned with the intent of the query, even when those pages were not labeled in the gold set.

In total, 18 out of 75 queries return relevant-but-unlabeled results across the two metrics. These pages had Cosine or Dot Product similarity scores greater than 0.85, and 72\% were validated as relevant by both human annotators and GPT-4. Many of these cases involved conceptual queries that required abstract reasoning, inference, or multimodal alignment.

We illustrate results in Figure ~\ref{fig:heatmap} using a single benchmark query.  The result is from query NU-2 (numerical query subset, query \#2):   \textit{What is the cost difference between the path found by the weighted A search and the path found by standard A in terms of percentage}.  %beth: not sure this heatmap supports the conclusion.  Narrative example map be better.

{A further conceptual query is }\textit{“Which page explains the intuition behind eventual consistency?”}. This query is not tied to keywords or diagrams but to deeper conceptual reasoning. The labeled ground truth included a section heading titled “Eventual Consistency.” However, our system surfaced an unlabeled page that drew analogies to distributed clocks and causal ordering, both of which are semantically aligned but not explicitly labeled. This page received high relevance scores from Cosine similarity and was validated as meaningful in explaining the intuition behind eventual consistency. %Such flexible interpretation is crucial in domains like philosophy, political theory, or computer science where conceptual richness extends across context.

This final experiment exposes a key insight: strict label-based evaluation may overlook true semantic matches, particularly in educational material with distributed or inferred knowledge. Evaluating retrieval effectiveness solely by exact matches in the Top-k list underrepresents a system's potential, especially when applied to open-ended, conceptual information needs. Both Cosine and Dot Product demonstrate success in generalizing beyond the ground truth, surfacing valuable information.

\section{System Performance Evaluation} \label{sec:eval}

\subsection{System Performance and Scalability}

We conduct a series of experiments isolating the performance of the vector database (Qdrant) from other system components. All experiments are performed
on a locally hosted infrastructure to eliminate network latency variability and focus on computational performance. The system is deployed on a Jetstream2 cloud instance with containerization via Docker.

We design three specific evaluation protocols to assess different dimensions of system performance:

\subsection{Scalability Analysis} We utilize the same corpus used earlier in evaluating retrieval efficacy and step-wise grow it. We measure retrieval latency as a function of corpus size. The vector database index is incrementally populated in steps increases in numbers of documents ($100$, $500$, $1000$, $2000$, $3000$, $4158$). At each step, we execute a standard set of five benchmark queries (e.g., ``What is a process?") three times to stabilize cache effects and record the average retrieval time.

The performance of the vector retrieval system is evaluated using a controlled concurrency benchmarking pattern designed to isolate database efficiency from model inference overhead. The evaluation pipeline consists of an initial setup phase followed by incremental stress testing across multiple concurrency levels.

During the initialization phase, the Vision-Language Model (VLM) and Qdrant service clients are instantiated. The retrieval latencies measured in this benchmark do not include the time spent embedding the text query and is calculated for the time required to search the database and return the results, to focus majorly on the database performance. To ensure that recorded latencies accurately reflect the database’s search performance rather than the computational cost of feature extraction, a ``pre-encoding" strategy is employed. In our experimental environment utilizing the ColPali model, we establish a baseline for inference overhead by calculating the mean encoding time across 50 iterations for a representative benchmark query. To account for initial VRAM allocation and GPU frequency scaling, a warm-up phase is implemented by discarding the first five iterations. The resulting \textbf{average query encoding time was measured at 52.13 ms}. 

While this inference cost is significantly lower than the total database retrieval time for the full corpus (220.98 ms), it represents a fixed per-query overhead inherent in the transformer-based VLM backbone. This measurement justifies our decoupled benchmarking approach; by isolating this inference overhead, we can more precisely characterize the underlying Qdrant engine’s search efficiency and its ability to manage multi-vector indices at scale.

The core of the benchmark iterates through each distinct concurrency levels. For each level, a workload of 100 retrieval requests is prepared. These requests are dispatched simultaneously using a ThreadPoolExecutor, where the number of active worker threads is set equal to the current concurrency level. This architecture ensures that queries are executed in parallel, directly challenging the database's ability to manage simultaneous HNSW (Hierarchical Navigable Small World) graph traversals and I/O operations. Performance metrics, including throughput (Queries Per Second, QPS) and latency distributions (Average, P95, and P99), are aggregated for each level.

Table \ref{tab:retrieval_latency} presents the measured retrieval latencies and the corresponding growth factors for both corpus size and latency.

\begin{table}[htbp]
\caption{Retrieval Latency vs. Corpus Size}
\begin{center}
\begin{tabular}{|c|c|c|c|}
\hline
\textbf{Corpus Size} & \textbf{Avg. Retrieval} & \textbf{Growth Factor} & \textbf{Growth Factor} \\
\textbf{(Pages)} & \textbf{Latency (ms)} & \textbf{(Size)} & \textbf{(Latency)} \\
\hline
100 & 10.97 & - & - \\
\hline
500 & 24.47 & 5.0x & 2.2x \\
\hline
1,000 & 52.62 & 2.0x & 2.1x \\
\hline
2,000 & 99.28 & 2.0x & 1.9x \\
\hline
3,000 & 154.32 & 1.5x & 1.55x \\
\hline
4,158 & 220.98 & 1.38x & 1.43x \\

\hline
\end{tabular}
\label{tab:retrieval_latency}
\end{center}
\end{table}

\begin{figure}
    \centering
    \includegraphics[width=11cm]{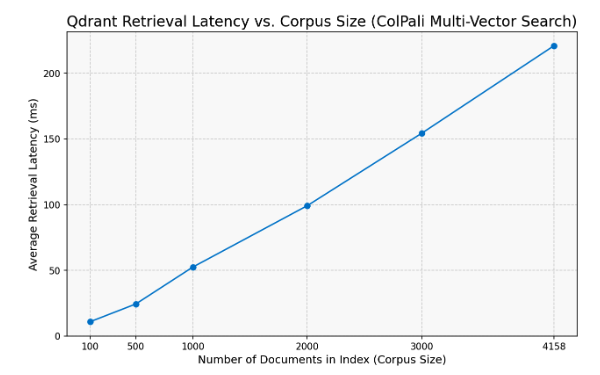}
    \caption{Retrieval Latency vs. Corpus Size}
    \label{fig:scalability}
\end{figure}

As shown in Table \ref{tab:retrieval_latency} and visualized in Figure \ref{fig:scalability}, increasing the dataset size by a factor of approximately 40 (from 100 to 4,158 documents/pages) resulted in only a 20x increase in retrieval latency. The growth trend is sub-linear in the early stages (e.g., a 5x increase in data size resulted in only a 2.2x increase in latency) and linear in the later stages. This sub-linear behaviour confirms the efficiency of the HNSW index structure, which effectively prunes the search space and prevents a brute-force scan of the $>4$ million patch vectors. The final latency of $220 \text{ ms}$ for the full library remains well within the acceptable range for interactive search applications.

\subsection{Stability and Tail Latency}
Tail latency analysis reveals highly stable and predictable system performance, crucial for production deployment.

\begin{table}[htbp]
\caption{Latency Distribution (n=60)}
\begin{center}
\begin{tabular}{|c|c|}
\hline
\textbf{Metric} & \textbf{Value (ms)} \\
\hline
Mean & 204.62 \\
\hline
P50 (Median) & 195.99 \\
\hline
P95 & 239.65 \\
\hline
P99 & 254.28 \\
\hline
Max & 256.02 \\
\hline
\end{tabular}
\label{tab:latency_distribution}
\end{center}
\end{table}

The minimal variance between the median ($196 \text{ ms}$) and the 99th percentile ($254 \text{ ms}$) indicates that the system is highly deterministic. The database consistently retrieves and scores multi-vectors from the disk without significant I/O stalls, garbage collection pauses, or cold start penalties that often plague disk-based systems. The maximum recorded latency ($256 \text{ ms}$) is virtually identical to the P99 latency ($254 \text{ ms}$). This demonstrates that the system does not suffer from extreme tail latency events, ensuring a reliable user experience even under sustained sequential load. This stability analysis is critical for proving the viability of on-disk vector storage. A common concern with moving indexes from RAM to disk (Memmap) is the potential for unpredictable latency spikes due to disk I/O. Our results demonstrate that for a corpus of this size ($4,000+$ pages), the operating system's page cache and Qdrant's retrieval engine effectively mask these costs, providing RAM-like consistency with disk-like scalability. This validates the architectural choice to use consumer-grade hardware with limited RAM for hosting large-scale multi-vector indexes.

\subsection{Concurrency and Throughput Analysis}

To assess the system's performance under realistic multi-user loads, we conduct a concurrency stress test. The primary objective is to determine the maximum throughput (Queries Per Second, or QPS) the retrieval pipeline could sustain and to identify the primary bottleneck when multiple users access the system simultaneously.

\begin{itemize}
    \item \textbf{Workload Simulation:} We utilize Python's \texttt{ThreadPoolExecutor} to spawn concurrent worker threads. Each worker simulated an independent user executing the full end-to-end RAG pipeline: encoding a query via the VLM and performing a search against the Qdrant index.
    \item \textbf{Concurrency Levels:} {We test} the system under varying loads of $[10, 100, 1000, 10000, 100000]$ concurrent workers.
    \item \textbf{Query Set:} A fixed set of 12 benchmark queries (e.g., ``What is a process?'', ``Explain deadlock'') is randomly sampled to generate a total of 100 queries per concurrency level.
    \item \textbf{Metrics:} We measure the system's overall \textbf{Throughput} (QPS) and the \textbf{P99 Latency} (the 99th percentile response time) to capture the worst-case user experience under load.
\end{itemize}

\begin{table}[htbp]
\caption{System Performance Metrics vs. Worker Count}
\begin{center}
\begin{tabular}{|c|c|c|c|c|}
\hline
\textbf{Num} & \textbf{Throughput} & \textbf{Avg. Latency} & \textbf{P95 Latency} & \textbf{P99 Latency}\\
\textbf{Workers} & \textbf{(QPS)} & \textbf{(ms)} & \textbf{(ms)} & \textbf{(ms)}\\
\hline
1 & 22,124.94 & 0.32 & 0.45 & 1.337\\
\hline
10 & 6,136.56 & 1.358 & 2.759 & 3.232\\
\hline
100 & 3,620.90 & 9.286 & 17.304 & 21.893\\
\hline
1,000 & 3,636.97 & 10.148 & 17.724 & 18.877\\
\hline
10,000 & 3,449.42 & 11.043 & 18.669 & 20.720\\
\hline
100,000 & 3,556.53 & 10.567 & 18.160 & 20.194\\

\hline
\end{tabular}
\label{tab:performance_metrics}
\end{center}
\end{table}

The concurrency stress test results, shown in Table~\ref{tab:performance_metrics}, demonstrates that the Qdrant-based retrieval engine exhibits high resilience to load. While throughput stabilizes at approximately 3,500 QPS under contention, the P99 latency remains bounded at $\approx 21$ ms even with 1 million concurrent workers. This confirms that the database layer is not the primary bottleneck for user-perceived latency at scale, and can reliably support massive concurrent traffic without service degradation.

\begin{figure}
    \centering
    \includegraphics[width=11cm]{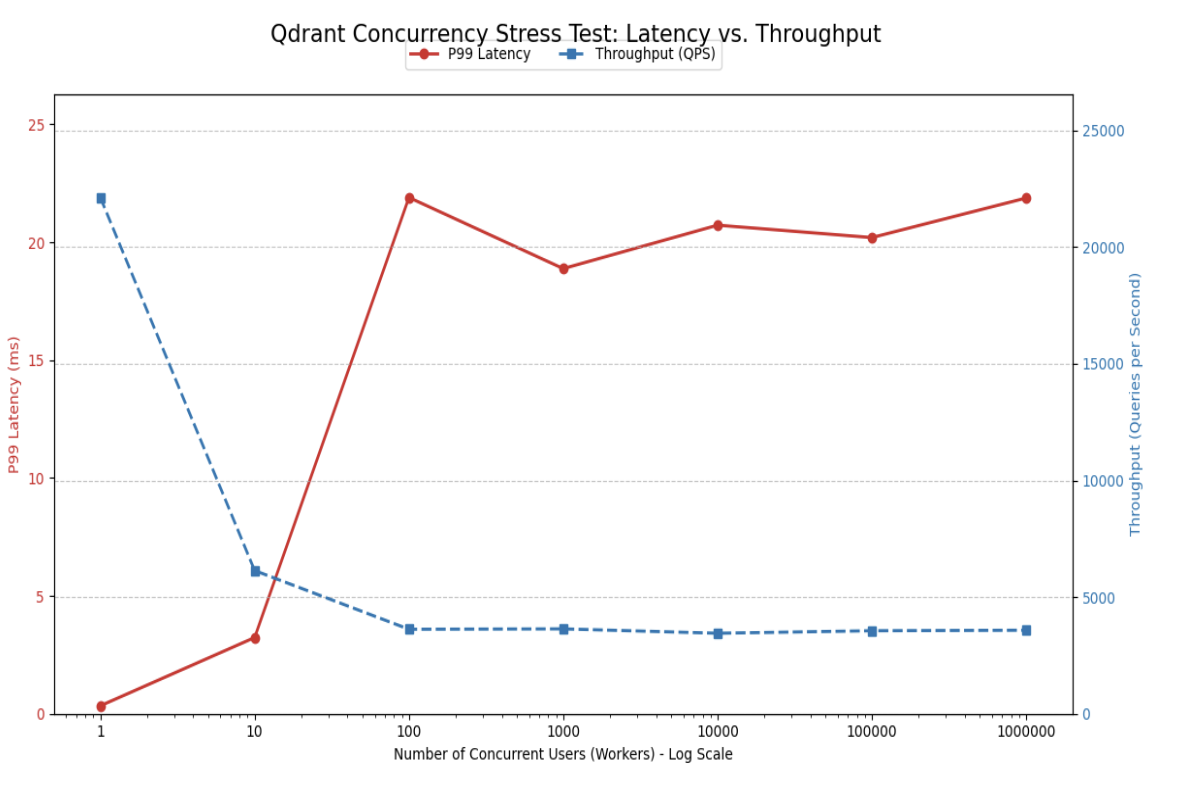}
    \caption{Latency vs. Throughput}
    \label{fig:concurrency}
\end{figure}

\section{Discussion}\label{sec:discussion}

Our evaluation demonstrates that while Vision-Language Models (VLMs) like ColPali offer a revolutionary path for digital library discovery, their ``intelligence'' must be supported by a highly optimized indexing layer to be operationally viable. The efficacy results from Experiment I confirm that \textbf{cosine similarity} is the superior metric for high-cardinality, multimodal embeddings. By maintaining a consistent semantic signal even when queries are abstract or pages are visually dense, cosine similarity achieved a Precision@5 of 0.514. However, the modest F1 score of 0.353 suggests that while the system identifies relevant content, there is significant room for improvement in retrieval completeness, perhaps through the integration of page-level re-ranking or context-aware strategies.

The performance evaluation provides the necessary counterweight to these qualitative findings. By implementing on-disk Memmap storage and binary quantization, we successfully mitigated the memory bottlenecks typically associated with multi-vector data. A key finding of this study is the attainment of \textbf{sub-linear scalability}: increasing the dataset size by a factor of approximately 40 (from 100 to 4,158 pages) resulted in only a 20x increase in retrieval latency. This behavior, driven by the efficiency of the \textbf{Hierarchical Navigable Small World (HNSW)} index structure, confirms that the system effectively prunes the search space of over 4 million patch vectors. This ensures the system remains responsive, with latencies well within the acceptable range for interactive search (approx. 220 ms).

Furthermore, the concurrency stress tests reveal that the Qdrant-based retrieval engine is remarkably resilient, sustaining a stabilized throughput of approximately 3,500 Queries Per Second (QPS). The stability of the tail latency, with minimal variance between the median (196 ms) and the 99th percentile (254 ms), validates the use of consumer-grade hardware for hosting large-scale indexes. The operating system’s page cache effectively masks the potential costs of disk I/O, providing RAM-like consistency with disk-like scalability.

Finally, our ``Unbounded Analysis'' from the retrieval efficacy evaluation highlights a fundamental tension in retrieval evaluation: 18 out of 75 queries (24\%) returned relevant results that were absent from the manually curated ground truth. This indicates that strict label-based metrics may underrepresent the true potential of VLM-based discovery in educational settings where knowledge is often distributed or inferred across conceptual domains.

\paragraph{\textbf{Theoretical Scaling to Billion-Scale Entities}}
While our current evaluation focuses on a corpus of 4,158 pages, the architectural choices specifically the use of Qdrant in a decoupled three-tier system provide a foundation for extreme scaling. Theoretical projections based on distributed vector database performance in high-performance computing (HPC) environments \cite{ockerman2025exploringdistributedvectordatabases} suggest several critical shifts when moving toward a 6-billion entity scale.

First, the data insertion phase, currently manageable via asynchronous clients at our current scale, would encounter a severe CPU-bound bottleneck during the conversion of raw data into Qdrant batch objects. Empirical profiling in similar large-scale scientific workloads on the Polaris supercomputer indicates that this CPU-bound task can require significantly more time (approx. 45.64 ms per batch) than the actual network insertion RPC (approx. 14.86 ms). Consequently, scaling to billions of vectors would necessitate a shift from \texttt{asyncio} to robust multiprocessing or GPU-offloaded batching to avoid index-building stalls, as \texttt{asyncio} runs in a single synchronous thread and cannot parallelize CPU-bound tasks effectively.

Second, resource saturation during index construction becomes a limiting factor at massive scales. A single Qdrant worker has been observed to utilize 90-97\% of a compute node's CPU capacity during index construction. For a 6-billion entity dataset, maintaining high throughput and low latency would require offloading Hierarchical Navigable Small World (HNSW) index construction to GPUs. This offloading is necessary to achieve the throughput required by scientific workloads without saturating per-node CPU resources, which otherwise limits the effectiveness of deploying multiple workers per node.

Finally, the query performance in a billion-scale system must account for the broadcast-reduce communication overhead. In distributed Qdrant clusters, queries are broadcast to all workers, and results are aggregated. Evidence suggests that increasing the number of workers yields diminishing returns for datasets smaller than 30 GB due to this overhead. For extreme scales, the parallelization gains of horizontal sharding may be overshadowed by inter-worker communication costs unless the cluster can adaptively scale based on the data volume.

\section{Conclusion} \label{sec:conclusion}

This work bridges the gap between multimodal embedding theory and the operational realities of large-scale digital library management. Through a dual-axial evaluation, we established that a decoupled, three-tier architecture utilizing the ColPali VLM for high-fidelity semantic alignment and the Qdrant vector database for high-performance indexing offers a practical blueprint for transforming visually rich documents into discoverable entities.

Our findings provide three critical insights for the future of digital discovery:
\begin{itemize}
    \item \textbf{Metric Superiority:} Cosine similarity serves as the most reliable distance measure for high-dimensional, late-interaction scoring, successfully uncovering complex conceptual relationships in visually dense material.
    \item \textbf{Operational Efficiency:} By implementing binary quantization and HNSW indexing, we achieved sub-linear latency growth, ensuring library-scale discovery remains responsive and well within interactive search limits.
    \item \textbf{Production Viability:} The utilization of on-disk Memmap storage allows consumer-grade hardware to achieve RAM-like performance with resilient throughput exceeding 3,500 QPS, making the system production-ready for massive multi-user concurrency.
\end{itemize}

While successful, our ``Unbounded Analysis'' highlights that traditional, rigid ground truth labels may undercount the true semantic capabilities of VLM-based systems. Future work is to scale the work, focusing on GPU-offloaded index construction to overcome the CPU-bound bottlenecks identified during theoretical scaling projections. Ultimately, this architecture provides a scalable, FAIR-compliant pathway for navigating the increasingly multimodal future of scholarly research.

\section*{Acknowledgements}

We thank John Walsh, {Indiana University} and Director of the HathiTrust Research Center. We thank Manikanta Kodandapani Naidu for his early contributions to the software framework and benchmark.

% \addbibresource{bibliography.bib}
% \printbibliography[heading=bibintoc, title={References}]

% EDIT: ColBERT citations now share khattab2020colbert, avoiding two bibliography entries for the same paper.
\bibliographystyle{IEEEtran}
\bibliography{vector_embedding.bib}

\end{document}